\documentclass[a4paper,12pt]{article}
\usepackage[dvips]{graphicx}
\usepackage[centertags]{amsmath}
\usepackage{amsfonts}
\usepackage{amssymb}
\usepackage{latexsym}
\usepackage{mathrsfs}
\usepackage{cite}

\RequirePackage{ifpdf} 
\ifpdf
  \RequirePackage[%
    unicode,%
    pdfstartview=FitH,%
    colorlinks,%
    hyperfootnotes=true,%
    bookmarksopen,%
    bookmarksnumbered,%
    bookmarksopenlevel=3%
  ]{hyperref}
  \RequirePackage{thumbpdf}
\else
  \RequirePackage{hyperref}
\fi \RequirePackage{hyperref}

\newcommand{\email}[1]{\href{mailto:#1}{\texttt{#1}}}

\newcommand{\braket}[2]{\ensuremath{\left\langle{#1}\!%
\mathrel{\left|{\vphantom{{#1} {#2}}}\right.%
\kern-\nulldelimiterspace}\!{#2}\right\rangle}}

\newcommand{\BK}[3]{\ensuremath{\left\langle{#1}\!%
\mathrel{\left|\vphantom{{#1}}{#2}\vphantom{{#3}}\right|%
\kern-\nulldelimiterspace}\!{#3}\right\rangle}}

\newcommand{\commute}[2]{\ensuremath{\left[{#1}\!%
\mathrel{\vphantom{{#1}},\vphantom{{#2}}%
\kern-\nulldelimiterspace}\!{#2}\right]}}

\newcommand{\uI}{\ensuremath{\mathrm{i}}}
\newcommand{\uE}{\ensuremath{\mathrm{e}}}
\newcommand{\uD}{\ensuremath{\mathrm{d}}}

\renewcommand{\vec}[1]{\ensuremath{\boldsymbol{\mathrm{#1}}}}

\makeatletter
\let\@afterindentfalse\@afterindenttrue
\@afterindenttrue \makeatother

\begin{document}



\title{\textbf{The Graviton Propagator with a Non-Conserved External Generating Source}}

\author{\textbf{E.~B.~Manoukian}\footnote{E-mail:~\email{manoukian\_eb@hotmail.com}.} \,and \textbf{S.~Sukkhasena} \\
{School of Physics, \ Institute of Science,} \\
{Suranaree University of Technology,} \\
{Nakhon~Ratchasima, 30000, Thailand}}

\date{}

\maketitle

\begin{abstract}
A novel general expression is obtained for the graviton propagator
from Lagrangian field theory by taking into account the necessary
fact that in the functional differential approach of quantum field
theory, in order to generate non-linearities in gravitation and
interactions with matter, the external source $T_{\mu\nu}$,
coupled to the gravitational field, should \textit{a priori} not
be conserved $\partial^\mu T_{\mu\nu}\neq 0$, so variations with
respect to its ten components may be varied
\textit{independently}. The resulting propagator is the one which
arises in the functional approach and does \textit{not} coincide
with the corresponding time-ordered product of two fields and it
includes so-called Schwinger terms. The quantization is carried
out in a gauge corresponding to physical states with two
polarization states to ensure positivity in quantum
applications.\\

\noindent\textbf{KEY~WORDS:} Graviton propagator; quantum gravity; non-conserved external sources; Schwinger terms 
\end{abstract}
\section{Introduction}\label{Section1}
A basic ingredient in quantum gravity computations is the graviton
propagator ([cf.1-5]). 
 The latter mediates the gravitational interaction between all
 particles to the leading order in the gravitational coupling
 constant. In the so-called functional differential treatment
 \cite{Manoukian_1986,Limboonsong_2006,Schwinger_1951,Manoukian_1985,Manoukian_2006}
  of quantum field theory, referred as the quantum dynamical
 principle approach, based on functional derivative techniques
 with respect to external sources coupled to the underlying fields
 in a theory, functional derivatives are taken of the so-called
 vacuum-to-vacuum transition amplitude. The latter generates
 \textit{n}-point functions by functional differentiations leading finally
 to transition amplitudes for various physical processes. For
 higher spin fields such as the electromagnetic vector potential
 $A^\mu$, the gluon field $A^\mu_a$, and certainly the
 gravitational field $h^{\mu\nu}$, the respective external sources
 $J_\mu$, $J_\mu^a$, $T_{\mu\nu}$, coupled to these fields, cannot
 \textit{a priori} taken to be conserved so that their respective
 components may be varied \textit{independently}. The consequences
 of relaxing the conservation of these external sources are highly
 non-trivial. For one thing the corresponding field propagators
 become modified. Also they have led to the rediscovery
 \cite{Manoukian_1986, Limboonsong_2006} of Faddeev--Popov \cite{Faddeev_1967}
 factors in non-abelian gauge theories and the discovery
 \cite{Limboonsong_2006} of even more generalized such factors,
 directly from the functional \textit{differential} treatment,
 via the application of the quantum dynamical principle, in the
 presence of external sources, without using commutation rules,
 and without even going to the well known complicated structures
 of the underlying Hamiltonians. A brief account of this is given
 in the concluding section for the convenience of the reader.

 For higher spin fields, the propagator and the time-ordered
 product of two fields do \textit{not} coincide as the former
 includes so-called Schwinger terms which, in general, lead to a
 simplification of the expression for the propagator over the
 time-ordered one. This is well known for spin 1 , and, as shown
 below, is also true for the graviton propagator. Let $h^{\mu\nu}$
 denote the gravitational field (see Sect.\ref{Section2}).
 We work in a gauge
\begin{equation}\label{Eqn1}
\partial_ih^{i\nu} = 0
\end{equation}
where $i=1,2,3;~\nu=0,1,2,3$, which guarantees that only two
states of polarization occur with the massless particle and
ensures positivity in quantum  applications avoiding non-physical
states. Let $T_{\mu\nu}$ denote an external source coupled to the
gravitational field $h^{\mu\nu}$ (see Sect.2), and let
$\braket{0_+}{0_-}^T$ denote the vacuum-to-vacuum transition
amplitude in the presence of the external source. The propagator
of the gravitational field is then defined by
\begin{align}\label{Eqn2}
\Delta_+^{\mu\nu;\sigma\lambda}(x,x') = \uI\left(
(-\uI)\frac{\delta}{\delta T_{\mu\nu }(x)}
(-\uI)\frac{\delta}{\delta T_{\sigma\lambda }(x')}
\braket{0_+}{0_-}^T\right)\Big/ \braket{0_+}{0_-}^T,
\end{align}
in the limit of the vanishing of the external source $T_{\mu\nu}$.
In more detail we may rewrite (\ref{Eqn2}) as
\begin{align}\label{Eqn3}
\Delta_+^{\mu\nu;\sigma\lambda}(x,x') = \uI
\frac{\BK{0_+}{\left(h^{\mu\nu}(x)h^{\sigma\lambda}(x')\right)_+}{0_-}^T}{\braket{0_+}{0_-}^T}
+ \frac{\BK{0_+}{\dfrac{\delta}{\delta T_{\mu\nu
}(x)}h^{\sigma\lambda }(x')}{0_-}^T}{\braket{0_+}{0_-}^T}
\end{align}
in the limit of vanishing $T_{\mu\nu}$, where the first term on
the right-hand side, up to the $\uI$ factor, denotes the
time-ordered product. In the second term, the functional
derivative with respect to the external source $T_{\mu\nu}(x)$ is
taken by keeping the independent field components of
$h^{\sigma\lambda}(x')$ fixed. The dependent field components
depend on the external source and lead to extra terms on the
right-hand side of (\ref{Eqn3}) in addition to the time-ordered
product and may be referred to as Schwinger terms. For a detailed
derivation of the general identity in (\ref{Eqn3}) see
Ref.\cite{Manoukian_2007} (see also \cite{Manoukian_2006}). These
additional terms lead to a simplification of the expression for
the propagator over the time-ordered product. Accordingly, the
propagator and the time-ordered product do \textit{not} coincide
and it is the propagator $\Delta_+^{\mu\nu;\sigma\lambda}$ that
appears in the functional approach and not the time-ordered
product. The derivation of the explicit expression for
$\Delta_+^{\mu\nu;\sigma\lambda}(x,x')$ follows by relaxing the
conservation of $T_{\mu\nu}$ and it includes 30 terms in contrast
to the well known case involving only 3 terms when a conservation
law of $T_{\mu\nu}$ is imposed. It is important to emphasize that
our interest here is in the propagator, the basic component which
appears in the theory, and not the time-ordered product. In the
concluding section, some additional pertinent comments are made
regarding our expression for the propagator. Our notation for the
Minkowski meter is $g^{\mu\nu} = \textrm{diag}[-1,1,1,1]$, also
quite generally we set $i,j,k,l = 1,2,3$, $a,b = 1,2$, while
$\mu,\nu,\sigma,\lambda= 0,1,2,3$.

\setcounter{section}{1}

\section{The Graviton Propagator}\label{Section2}
For the Lagrangian density of the gravitational field $h^{\mu\nu}$
coupled to an external source $T_{\mu\nu}$, we take
\begin{align}\label{Eqn4}
\mathscr{L} = -\frac{1}{2}&\partial^\alpha
h^{\mu\nu}\partial_\alpha h_{\mu\nu} + \frac{1}{2}\partial^\alpha
h^\sigma{}_\sigma \partial_\alpha h^\beta{}_\beta -
\partial^\alpha h_{\alpha\mu} \partial^\mu h^\sigma{}_\sigma \nonumber \\+ &
\frac{1}{2}\partial_\alpha h^{\alpha\nu}\partial^\beta
h_{\beta\nu} + \frac{1}{2}\partial_\alpha h^\mu{}_\nu \partial_\mu
h^{\alpha\nu} + h^{\mu\nu} T_{\mu\nu} ,
\end{align}
where $h^{\mu\nu}=h^{\nu\mu}$, and as a result $T_{\mu\nu}$ is
chosen to be symmetric. We consider the ten components of
$T_{\mu\nu}$ to be independent by, \textit{a priori}, not imposing
a conservation law for $T_{\mu\nu}$. The action corresponding to
the Lagrangian density in (\ref{Eqn4}), in the absence of the
external source $T_{\mu\nu}$, is invariant under the gauge
transformation $h^{\mu\nu} \to h^{\mu\nu} + \partial^\mu\xi^\nu +
\partial^\nu\xi^\mu + \partial^\mu\partial^\nu\xi$.
The gauge constraint in (\ref{Eqn1}) allows us to solve, say,
$h_{3\nu}$, in terms of other components:
\begin{align}\label{Eqn5}
h_{30} = -(\partial_3)^{-1}\partial_a h_{a0},\, h_{3a} =
-(\partial_3)^{-1}\partial_b h_{ba}, \nonumber \\ h_{33} =
-(\partial_3)^{-1}\partial_a h_{3a} = (\partial_3)^{-2}\partial_a
\partial_b h_{ab},
\end{align}
where $a,b = 1,2$. Upon substituting the expressions for
$h_{3\nu}$ in (\ref{Eqn4}), and varying $h_{ab}$, we obtain
\begin{align}\label{Eqn6}
(\square & h_{ab} + T_{ab}) - \frac{\partial_b}{\partial_3}
(\square h_{a3} + T_{a3}) - \frac{\partial_a}{\partial_3}(\square
h_{b3}+T_{b3}) \nonumber \\
&+ \frac{\partial_a\partial_b}{(\partial_3)^2}(\square h_{33} +
T_{33}) + \left[\delta_{ab}+
\frac{\partial_a\partial_b}{(\partial_3)^2}
\right](\vec{\partial}^2 h_{00} - \square h_{ii}) = 0,
\end{align}
$a,b = 1,2$. Upon multiplying (\ref{Eqn6}) by $(\delta_{ab} -
\partial_a\partial_b / \vec{\partial}^2)$, where $\vec{\partial}^2 =
\partial^i\partial_i$, $i=1,2,3$, some tedious algebra leads to
\begin{align}\label{Eqn7}
-\vec{\partial}^2 h_{00} = -\frac{1}{2} \square h_{ii} +
\frac{1}{2}\left(\delta_{ij} -
\frac{\partial_i\partial_j}{\vec{\partial}^2}\right)T_{ij}.
\end{align}

On the other hand, with the expressions for $h_{3\nu}$ in
(\ref{Eqn5}) replaced in (\ref{Eqn4}), variations with respect to
$h_{00}$, $h_{0a}$, $a=1,2$, give, respectively,
\begin{align}\label{Eqn8}
-\vec{\partial}^2 h_{ii} = T_{00},
\end{align}
\begin{align}\label{Eqn9}
-\vec{\partial}^2 h_{0a}  +
\frac{\partial_a}{\partial_3}\vec{\partial}^2 h_{03} =
\left(T_{0a} - \frac{\partial_a}{\partial_3}T_{03}\right)
\end{align}

We note that (\ref{Eqn9}) is valid if we formally replace $a$ by 3
since this simply gives $0=0$. Accordingly, we may rewrite
(\ref{Eqn9}) as
\begin{align}\label{Eqn10}
-\vec{\partial}^2 h_{0i} +
\frac{\partial_i}{\partial_3}\vec{\partial}^2 h_{03} = T_{0i} -
\frac{\partial_i}{\partial_3}T_{03}
\end{align}
where $i = 1, 2, 3$. Upon taking the divergence $\partial^i$ of
(\ref{Eqn10}) and using (\ref{Eqn1}), we obtain
\begin{align}\label{Eqn11}
\frac{\partial_i}{\partial_3}\vec{\partial}^2 h_{03} =
\frac{\partial_i}{\vec{\partial}^2} \left(\partial_j T_{0j} -
\frac{\vec{\partial}^2}{\partial_3}T_{03}\right),
\end{align}
which upon substitution in (\ref{Eqn10}) gives
\begin{align}\label{Eqn12}
-\vec{\partial}^2 h_{0i} = \left(\delta_{ij} -
\frac{\partial_i\partial_j}{\vec{\partial}^2}\right) T_{0j}.
\end{align}

Also upon substitution (\ref{Eqn8}) in (\ref{Eqn7}), and using the
fact that $\square = \vec{\partial}^2 - \partial^2_0$, we obtain
for (\ref{Eqn7})
\begin{align}\label{Eqn13}
-\vec{\partial}^2 h_{00} = T_{00} + \frac{T}{2} -
\frac{1}{2\vec{\partial}^2}\left(\partial^0\partial^0 T_{00} +
\partial_i\partial_j T_{ij}\right),
\end{align}
where $T=g^{\mu\nu}T_{\mu\nu} = T^\nu {}_\nu$.

Equations (\ref{Eqn8}), (\ref{Eqn12}), (\ref{Eqn13}) are not
equations of motion as they involve no time derivatives of the
corresponding fields and they yield to constraints which together
the gauge condition in (\ref{Eqn1}) give rise to two degrees of
freedom corresponding to two polarization states for the graviton
as it should be.

We now substitute the expression for $-\vec{\partial}^2h_{00}$, as
given in (\ref{Eqn13}), in (\ref{Eqn6}) and use (\ref{Eqn8}) to
obtain an equation involving $h_{ij}$, $i,j=1,2,3$. Upon
multiplying the resulting equation from (\ref{Eqn6})  by
$\partial_a\partial_b$ and using the expressions for $h_{33}$ in
(\ref{Eqn5}) we obtain after some very tedious algebra
\begin{align}\label{Eqn14}
(\square h_{33} + T_{33}) - &\frac{1}{2}\left(1 -
\frac{(\partial_3)^2}{\vec{\partial}^2}\right)T +
\frac{1}{2\vec{\partial}^2}\left(-\partial^0\partial^0 T_{00} +
\partial^i\partial^j T_{ij}\right)\nonumber\\
-&\frac{2}{\vec{\partial}^2}\partial^i\partial_3 T_{i3} +
\frac{(\partial_3)^2}{2\vec{\partial}^2}\left(\frac{\partial^i\partial^j}{\vec{\partial}^2}T_{ij}
+ \frac{\partial^0\partial^0}{\vec{\partial}^2}T_{00}\right)= 0.
\end{align}

Similarly, upon multiplying (\ref{Eqn6}) by $\partial_a$ and using
the expression for $h_{b3}$ in (\ref{Eqn5}), we obtain
\begin{align}\label{Eqn15}
(\square h_{b3} + T_{b3}) - \frac{1}{\vec{\partial}^2}
\left[\partial_3\partial^i T_{ib} + \partial_b\partial^i T_{i3} -
\frac{\partial_b\partial_3}{2}\left(\frac{\partial^i\partial^j
}{\vec{\partial}^2}T_{ij} + \frac{\partial^0\partial^0
}{\vec{\partial}^2}T_{00} + T\right)\right] = 0
\end{align}

To obtain the equation for $h_{ab}$, we substitute (\ref{Eqn14}),
(\ref{Eqn15}) in (\ref{Eqn6}), to obtain after some lengthy
algebra
\begin{align}\label{Eqn16}
&(\square h_{ab} + T_{ab}) - \frac{1}{2} \left(\delta_{ab} -
\frac{\partial_a\partial_b}{\vec{\partial}^2}\right)T  -
\frac{1}{\vec{\partial}^2}(\partial_a\partial^i T_{ib} +
\partial_b\partial^i T_{ia})\nonumber \\
&+\frac{\delta_{ab}}{2\vec{\partial}^2}\left(-\partial^0\partial^0
T_{00} + \partial^i\partial^j T_{ij}\right) +
\frac{\partial_a\partial_b}{2(\vec{\partial}^2)^2}\left(\partial^i\partial^j
T_{ij} + \partial^0\partial^0 T_{00}\right) = 0.
\end{align}

Equations (\ref{Eqn14}), (\ref{Eqn15}), (\ref{Eqn16}) may be now
combined in the form
\begin{align}\label{Eqn17}
-\square h_{ij} =  T_{ij} &- \frac{1}{2} \left(\delta_{ij} -
\frac{\partial_i\partial_j}{\vec{\partial}^2}\right) \left(T +
\frac{\partial^0\partial^0}{\vec{\partial}^2}T_{00}\right)\nonumber\\
&-\frac{1}{\vec{\partial}^2}\left[\partial_i\partial^k T_{kj} +
\partial_j\partial^k T_{ki} - \frac{1}{2}\left(\delta_{ij} + \frac{\partial_i\partial_j}{\vec{\partial}^2}\right)\partial^k\partial^l
T_{kl}\right],
\end{align}
where $i,j,k,l= 1,2,3$.

Equations (\ref{Eqn17}), (\ref{Eqn13}), (\ref{Eqn12}) give the
equations for the various components of $h_{\mu\nu}$. To obtain
the unifying equation for $h_{\mu\nu}$, we note that we may write
\begin{align}\label{Eqn18}
h^{\mu\nu} = g^{\mu i}h_{ij}g^{j\nu} + g^{\mu i}h_{i0}g^{0\nu} +
g^{\mu 0}h_{0j}g^{j\nu} + g^{\mu 0}h_{00}g^{0\nu},
\end{align}
with $i,j=1,2,3; \mu,\,\nu=0,1,2,3$, and use in the process the
identity
\begin{align}\label{Eqn19}
g^{\mu i}\partial_i = (\partial^\mu + N^\mu\partial_0),
\end{align}
where $N^\mu$ is the unit time-like vector ($N^\mu N_\mu=-1$)
\begin{align}\label{Eqn20}
(N^\mu)= (g^\mu{}_0) = (1,0,0,0).
\end{align}

Finally, we use the identity relating a tensor
$A_{\lambda\sigma}$, e.g., to the components $A_{ij}$ as follows:
\begin{align}\label{Eqn21}
g^{\mu i}A_{ij}g^{j\nu} = \left[g^{\mu\lambda}g^{\nu\sigma} +
N^\mu N^\lambda g^{\sigma\nu} + N^\nu N^\sigma g^{\lambda\mu}  +
N^\mu N^\nu N^\lambda N^\sigma\right]A_{\lambda\sigma},
\end{align}
and the fact that $\square= \vec{\partial}^2 - \partial^{0^2}$. A
lengthy analysis from (\ref{Eqn12}), (\ref{Eqn13}), (\ref{Eqn17})
then gives the following explicit expression for $h^{\mu\nu}$:
\begin{align}\label{Eqn22}
h^{\mu\nu} = \frac{1}{(-\square - \uI\epsilon)}&\Big\{
\frac{g^{\mu\lambda}g^{\nu\sigma} + g^{\mu\sigma}g^{\nu\lambda}
-g^{\mu\nu}g^{\sigma\lambda}}{2} \nonumber \\
&+ \frac{1}{2\vec{\partial}^2}
\Big[g^{\mu\nu}\partial^\sigma\partial^\lambda +
g^{\sigma\lambda}\partial^\mu\partial^\nu -
g^{\nu\sigma}\partial^\mu\partial^\lambda -
g^{\nu\lambda}\partial^\mu\partial^\sigma\nonumber\\
&-g^{\mu\sigma}\partial^\nu\partial^\lambda -
g^{\mu\lambda}\partial^\nu\partial^\sigma +
\frac{\partial^\mu\partial^\nu\partial^\sigma\partial^\lambda}{\vec{\partial}^2}\Big]
\nonumber \\
+\frac{1}{2}\left(g^{\mu\nu} +
\frac{\partial^\mu\partial^\nu}{\vec{\partial}^2}\right)&\left(\frac{N^\sigma\partial^\lambda
+ N^\lambda\partial^\sigma}{\vec{\partial}^2}\right)\partial_0 +
\frac{1}{2}\left(g^{\sigma\lambda}+\frac{\partial^\sigma\partial^\lambda}{\vec{\partial}^2}\right)\left(\frac{N^\nu\partial^\mu
+ N^\mu\partial^\nu}{\vec{\partial}^2}\right)\partial_0 \nonumber
\\
-\frac{1}{2}\Big[g^{\nu\sigma}&(N^\mu\partial^\lambda +
N^\lambda\partial^\mu) + g^{\nu\lambda}(N^\mu\partial^\sigma +
N^\sigma\partial^\mu) \nonumber\\
&+g^{\mu\sigma}(N^\nu\partial^\lambda + N^\lambda\partial^\nu) +
g^{\mu\lambda}(N^\nu\partial^\sigma +
N^\sigma\partial^\nu)\Big]\frac{\partial_0}{\vec{\partial}^2}
\nonumber \\
&+ \frac{\partial^\mu\partial^\nu}{\vec{\partial}^2}N^\sigma
N^\lambda +
\frac{\partial^\sigma\partial^\lambda}{\vec{\partial}^2}N^\mu
N^\nu \Big\}T_{\sigma\lambda} \nonumber\\
&+\frac{1}{\vec{\partial}^2}
\left\{\frac{\partial^\mu\partial^\nu}{\vec{\partial}^2}N^\sigma
N^\lambda +
\frac{\partial^\sigma\partial^\lambda}{\vec{\partial}^2}N^\mu
N^\nu\right\}T_{\sigma\lambda},
\end{align}

$\epsilon \to +0$.\\

From (\ref{Eqn22}) the explicit expression for the graviton
propagator $\Delta_+^{\mu\nu;\sigma\lambda}(x,x')$ emerges as:
\begin{align}\label{Eqn23}
\Delta_+^{\mu\nu;\sigma\lambda}(x,x')  = \int\dfrac{(\uD
k)}{(2\pi)^4}\uE^{\uI
k(x-x')}\left[\frac{\Delta_1^{\mu\nu;\sigma\lambda}(k)}{k^2-\uI\epsilon}
+ \frac{\Delta_2^{\mu\nu;\sigma\lambda}(k)}{\vec{k}^2}\right],
\end{align}
$\epsilon \to +0$, where $(\uD k)= \uD k^0\uD k^1\uD k^2\uD k^3$,
$k^2 = \vec{k}^2 - k^{0^2}$, and
\begin{align}\label{Eqn24}
\Delta_1^{\mu\nu;\lambda\sigma}(k) =
\frac{(g^{\mu\lambda}g^{\nu\sigma} + g^{\mu\sigma}g^{\nu\lambda} -
g^{\mu\nu}g^{\sigma\lambda})}{2}&\nonumber \\
        +\frac{1}{2\vec{k}^2}\Big[g^{\mu\nu}k^\sigma k^\lambda +
        g^{\sigma\lambda}k^\mu k^\nu &-g^{\nu\sigma}k^\mu k^\lambda
        -g^{\nu\lambda}k^\mu k^\sigma  \nonumber\\
-g^{\mu\sigma}k^\nu k^\lambda -g^{\mu\lambda}k^\nu k^\sigma &+
\frac{k^{\mu}k^{\nu}k^{\sigma}k^{\lambda}}{\vec{k}^2}\Big]
\nonumber\\
        -\frac{1}{2}\left(g^{\mu\nu} +
        \frac{k^{\mu}k^\nu}{\vec{k}^2}\right)\left(\frac{N^\sigma{}k^\lambda
        + N^\lambda{}k^\sigma}{\vec{k}^2}\right)k^0 & \nonumber\\
-\frac{1}{2}\left(g^{\sigma\lambda} +
\frac{k^\sigma{}k^\lambda}{\vec{k}^2}\right)\left(\frac{N^\nu{}k^\mu
+N^\mu{}k^\nu}{\vec{k}^2}\right)k^0&\nonumber \\
    +\frac{1}{2}\Big[g^{\nu\sigma}(N^\mu{}k^\lambda + N^\lambda{}k^\mu) + g^{\nu\lambda}(N^\mu{}k^\sigma + N^\sigma{}k^\mu)&\nonumber \\
        +g^{\mu\sigma}(N^\nu{}k^\lambda + N^\lambda{}k^\nu) +
        g^{\mu\lambda}(N^\nu{}k^\sigma &+
        N^\sigma{}k^\nu)\Big]\frac{k^0}{\vec{k}^2}\nonumber\\
+\frac{k^\mu k^\nu}{\vec{k}^2}N^\sigma N^\lambda + \frac{k^\sigma
k^\lambda}{\vec{k}^2} N^\mu N^\nu &,
\end{align}
\begin{align}\label{Eqn25}
\Delta_2^{\mu\nu;\lambda\sigma}(k)  =  \frac{k^\mu
k^\nu}{\vec{k}^2} N^\sigma N^\lambda + \frac{k^\sigma
k^\lambda}{\vec{k}^2} N^\mu N^\nu.
\end{align}

The vacuum-to-vacuum transition amplitude for the gravitational
field coupled to an external source is then given by
\begin{align}\label{Eqn26}
\braket{0_+}{0_-}^T  = \exp \left[\frac{\uI}{2}\int (\uD x)(\uD
x')
T_{\mu\nu}(x)\Delta_+^{\mu\nu;\sigma\lambda}(x,x')T_{\sigma\lambda}(x')\right].
\end{align}
with the graviton propagator given by the explicit expression in
(\ref{Eqn23}) - (\ref{Eqn25}). Now we are ready to make pertinent
comments concerning the graviton propagator thus obtained.

\section{Conclusion}\label{Section3}
We have derived a novel expression for the graviton propagator,
from Lagrangian field theory, valid for the case when the external
source $T_{\mu\nu}$ coupled to the gravitational field is not
necessarily conserved, by working in a gauge where only two
polarization physical states of the graviton arise to ensure
positivity in the quantum treatment thus avoiding non-physical
states. That such a conservation should \textit{a priori} not to
be imposed is a necessary mathematical requirement so that all the
ten components of the external source $T_{\mu\nu}$ may be varied
independently in order to generate interactions of the
gravitational field with matter and produce non-linearity of the
gravitational field itself in the functional procedure. The latter
requirement arises by noting that such interactions are generated
by the application [cf.6, 7] of some functional
$F[-\uI\delta/\delta T_{\mu\nu}]$ to $\braket{0_+}{0_-}^T$, where
$\braket{0_+}{0_-}$ corresponding to other particles, as well as
functional derivatives of their corresponding sources in $F$, have
been  suppressed to simplify the notation. Accordingly, to vary
the ten components of $T_{\mu\nu}$ independently, no conservation
may \textit{a priori} be imposed. The  $1/\vec{k^2}$ terms in
(\ref{Eqn23}) - (\ref{Eqn25}) are apparent singularities due to
the sufficient powers in $k$ in the corresponding denominators and
the three-dimensional character of space, in the same way that
this happens for the photon propagator in the Coulomb gauge in
quantum electrodynamics, and give rise to static $1/r$ type
interactions complicated by the tensorial character  of a spin two
object. It is important to note that for a conserved $T_{\mu\nu}$,
i.e., for $\partial^\mu T_{\mu\nu}= 0 $, all the terms in the
propagators in (\ref{Eqn23}), with the exception of the terms
$(g^{\mu\lambda}g^{\nu\sigma} + g^{\mu\sigma}g^{\nu\lambda} -
g^{{\mu\nu}}g^{\sigma\lambda})/2$, do not contribute in
(\ref{Eqn26}) since \textit{all} the other terms in (\ref{Eqn24}),
(\ref{Eqn25}) involve derivatives of $T_{\mu\nu}$ and the graviton
propagator $\Delta_+^{\mu\nu;\sigma\lambda}(x,x')$ effectively
\textit{goes over} to the well documented expression
\begin{align}\label{Eqn27}
\frac{1}{(-\square -
\uI\epsilon)}\frac{(g^{\mu\lambda}g^{\nu\sigma} +
g^{\mu\sigma}g^{\nu\lambda} - g^{{\mu\nu}}g^{\sigma\lambda})}{2},
\end{align}
which has been known for years [cf.1, 2]. This is unlike the
corresponding time-ordered product which does not go over to the
result in (\ref{Eqn27}) for $\partial^\mu T_{\mu\nu}= 0$  This may
be shown by solving for the time-ordered product in (\ref{Eqn3})
in terms of the propagator and carrying out explicitly, say, the
functional derivatives $\delta h^{0i}/\delta T_{\mu\nu}$, $\delta
h^{00}/\delta T_{\mu\nu}$, as arising on the right-hand side of
(\ref{Eqn3}), by using, in the process, Eqs.
(\ref{Eqn12}),(\ref{Eqn13}). In any case, it is the propagator
$\Delta_+^{\mu\nu;\sigma\lambda}$, as given in (\ref{Eqn23}), is
the one that appears in the theory and not the time-ordered
product as is often na\"{i}vely assumed. After all the functional
derivatives with respect to $T_{\mu\nu}$ are carried out in the
theory, one may impose a conservation law on $T_{\mu\nu}$ or even
set $T_{\mu\nu}$ equal to zero if required on physical grounds.
Such methods have led to the discovery
\cite{Manoukian_1986,Limboonsong_2006}, in the functional quantum
dynamical principle differential approach, of Faddeev--Popov (FP)
factors, and of their generalizations, in non-abelian gauge
theories such as in QCD and in other theories.

Re-iterating the discussion above, the relevance of the analysis
and the explicit expression derived for the graviton propagator
for, \textit{a priori}, not conserved external source $T_{\mu\nu}
:\partial^\mu T_{\mu\nu} \neq 0$ is immediate. If, in contrast, a
conservation law is \textit{a priori}, imposed then variations
with respect to one of the components of $T_{\mu\nu}$ would
automatically imply, via such a conservation law, variations with
respect some of its \textit{other} components as well. A problem
that may arise otherwise, may be readily seen from a simple
example. The functional derivative of an expression like
$[a_{\mu\nu}(x) + b(x)\partial_\mu\partial_\nu]T^{\mu\nu}(x)$,
with respect to a component $T^{\sigma\lambda}(x')$ is
$(1/2)[a_{\mu\nu}(x) +
b(x)\partial_\mu\partial_\nu](\delta_\sigma{}^\mu\delta_{\lambda}{}^\nu
+ \delta_{\lambda}{}^\mu\delta_\sigma{}^\nu)\delta^4(x,x')$, where
$a_{\mu\nu}(x)$, $b(x)$, for example, depend on $x$, and not
$(1/2)a_{\mu\nu}(x)(\delta_\sigma{}^\mu\delta_{\lambda}{}^\nu +
\delta_{\lambda}{}^\mu\delta_\sigma{}^\nu)\delta^4(x,x')$ as one
may na\"{i}vely assume by, \textit{a priori} imposing a
conservation law. Also, as mentioned above, the present method,
based on the functional differential treatment, as applied to
non-abelian gauge theories such as QCD
\cite{Manoukian_1986,Limboonsong_2006} leads automatically to the
presence of the FP determinant modifying na\"{i}ve Feynman rules.
The \textit{physical} relevance of such a factor is important as
its omission would lead to a violation of unitarity. For the
convenience of the reader we briefly review, before closing the
concluding section, on how the FP determinant arises in the
functional differential treatment
\cite{Manoukian_1986,Limboonsong_2006}.

Consider, for simplicity of the demonstration, the non-abelian
gauge theory with Lagrangian density
\begin{align}\label{Eqn28}
\mathscr{L} = -\frac{1}{4}G_{\mu\nu}^a G_a^{\mu\nu} +
J_a^{\mu}A_{\mu}^a
\end{align}
where $J_a^{\mu}$ is an external source taken, \textit{a priori},
not to be conserved. Here
\begin{align}\label{Eqn29}
G_{\mu\nu}^a = \partial_{\mu}A_{\nu}^a - \partial_{\nu}A_{\mu}^a +
g_of^{abc} A_{\mu}^b A_{\nu}^c
\end{align}
We work in the Coulomb gauge. The gauge field propagator, in
analogy to the graviton one in (\ref{Eqn24}), (\ref{Eqn25}), is
given by
\begin{align}\label{Eqn30}
D_{ab}^{\mu\nu} = \delta_{ab}[g^{\mu\nu} -
                    \frac{(\partial^{\mu}\partial^{\nu}
                        + N^{\mu}\partial^{\nu}\partial_0
                        + N^{\nu}\partial^{\mu}\partial_0)}
                    {\vec{\partial}^2}]
                    \frac{1}{-\Box-\uI\varepsilon}
\end{align}
with $k= 1, 2, 3$.

The quantum dynamical principle states that
\begin{align}\label{Eqn31}
\frac{\partial}{\partial g_o}\braket{0_+}{0_-} = \uI\BK{0_+}{\int
(\uD x)\frac{\partial}{\partial g_o}\mathscr{L}(x)}{0_-}
\end{align}
where, with $k= 1, 2, 3,$
\begin{align}\label{Eqn32}
\frac{\partial}{\partial g_o}\mathscr{L}(x) =
-f^{abc}A_{k}^{b}(A_0^c G_a^{k0} + \frac{1}{2}A_l^c G_a^{kl})
\end{align}
and $G_a^{kl}$ may be expressed in terms of independent fields,
that is, for which the canonical conjugate momenta do not vanish.
On the other hand, $G_a^{k0}$ depends on the dependent field
$A_a^0$. By using the identity
\begin{align}\label{Eqn33}
(-\uI)\frac{\delta}{\delta
J_a^{\mu}(x')}\BK{0_+}{\mathscr{O}}{0_-}
            =   \BK{0_+}{(A_{\mu}^a(x')\mathscr{O}(x))_+}{0_-}
            -\uI \BK{0_+}{\frac{\delta}{\delta J_a^{\mu}(x')}\mathscr{O}(x)}{0_-}
\end{align}
for an operator $\mathscr{O}(x)$, where $(...)_+$ denotes the
time-ordered product, and the functional derivative
$\delta\mathscr{O}(x)/\delta J_a^{\mu}(x')$ in the second term on
the right-hand side of (\ref{Eqn33}) is taken by keeping the
independent fields and their canonical conjugate kept fixed in
$\mathscr{O}(x)$, after the latter is expressed in terms of these
fields, together, possibly, in terms of the dependent fields and
the external current \cite{Limboonsong_2006, Manoukian_2007}.

From the Lagrangian density in (\ref{Eqn28}), the following
relation follows
\begin{align}\label{Eqn34}
G_a^{k0} = \pi_a^{k} - \partial^k D_{ab}J_b^0
\end{align}
as a matrix equation, where $\pi_a^k$ denotes the canonical
conjugate momentum of $A_a^k$, and $D_{ab}$ is the Green operator
satisfying
\begin{align}\label{Eqn35}
[\delta^{ac}\vec{\partial}^2
    + g_o f^{abc} A_k^b\partial^k]D^{cd}(x,x';g_o) =
    \delta^4(x,x')\delta^{ad}
\end{align}
Accordingly, with, \textit{a priori}, non-conserved $J_a^\mu(x')$,
we may vary each of its components independently to obtain from
(\ref{Eqn34})
\begin{align}\label{Eqn36}
\frac{\delta}{\delta J_a^\mu(x')}G_a^{k0}(x) = -\delta_\mu{}^0
\partial^k D_{ac}(x,x';g_o)
\end{align}

Hence from (\ref{Eqn32}), (\ref{Eqn33}), and (\ref{Eqn36}), we may
write
\begin{align}\label{Eqn37}
\langle 0_+|\frac{\partial}{\partial g_o}\mathscr{L}(x)|0_-\rangle
        = [(\frac{\partial}{\partial g_o}\mathscr{L})' + \uI f^{bca}A'{}_k^b\partial^k
        D'{}^{ac}(x,x;g_o)]\braket{0_+}{0_-}
\end{align}
where the primes mean to replace $A_{\mu}^c(x)$ in the
corresponding expressions by the functional differential operator
$(-\uI)\delta/\delta J_c^{\mu}(x)$.

Clearly, upon an elementary integration over $g_o$ in
(\ref{Eqn31}) by using, in the process, (\ref{Eqn37}) and the
equation for $D^{ac}$ in (\ref{Eqn35}), we \textit{obtain} the FP
determinant
\begin{align}
\exp \textrm{Tr}\ln [1-\uI g_o
\frac{1}{\vec{\partial}^2}A'_k\partial^k]
\end{align}
as a multiplicative modifying differential operating factor in
$\braket{0_+}{0_-}$. For additional related details see
\cite{Manoukian_1986, Limboonsong_2006} and also for further
generalizations of the occurrence of such factors in field theory.

It is interesting to extend such analyses \cite{Manoukian_1986,
Limboonsong_2006}, as well as of gauge transformations
\cite{Manoukian_1986}, and covariance \cite{Manoukian_1987}, to
theories involving gravity. This would be exponentially much
harder to do and will be attempted in further investigations. In
this regard, our ultimate interest is in aspects of
renormalizability \cite{Manoukian_1983} and rules for physical
applications that would follow from our, \textit{a priori},
systematic analysis carried out at the outset, in a quantum
setting with the newly modified propagator, by a functional
\textit{differential} treatment, in the presence of external
sources, to generate not-linearities in gravitation and
interactions with matter.
\\

\end{document}